\documentclass[10pt,english]{article}
\usepackage[T1]{fontenc}
\usepackage[latin1]{inputenc}
\usepackage{graphicx}
\usepackage{geometry}
\makeatletter

\providecommand{\LyX}{L\kern-.1667em\lower.25em\hbox{Y}\kern-.125emX\@}

 \newcommand{\lyxaddress}[1]{
   \par {\raggedright #1
   \vspace{1.4em}
   \noindent\par}
 }

\usepackage{babel}

\makeatother

\begin{document}

\title{Violation of the Fluctuation-Dissipation Theorem for fast superdiffusion}

\author{Ismael V. L. Costa, Rafael Morgado, Marcos V. B. T. Lima, Fernando
A. Oliveira}

\maketitle

\lyxaddress{Institute of Physics and International Center of Condensed Matter
Physics, University of Bras\'{\i}lia, CP 04513, 70919-970, Bras\'{\i}lia-DF,
Brazil}


\begin{abstract}
We study anomalous diffusion for one-dimensional systems described
by a generalized Langevin equation. We show that superdiffusion can
be classified in normal superdiffusion and fast superdiffusion. For
fast superdiffusion we prove that the Fluctuation-Dissipation Theorem
does not hold, which induces an effective temperature in the system.
This effective temperature is a signature of metastability found in
many complex system such as Spin-Glass and granular material.
\end{abstract}

\section{Introduction}

Since its formulation, the Fluctuation-Dissipation Theorem (FDT) has
played a central role\cite{Kubo,Kubo2} in non-equilibrium statistical
mechanics in the linear response regime (LNESM). It reaches such an
importance that a full formulation of LNESM is given \cite{Kubo2}
based on it. In the last 30 years, fundamental concepts and methods
have been developed \cite{Kubo,Kubo2,Mori,Evans,Lee2,Lee3} and a large number of
connections have been established. A necessary requirement for the
FDT is that the time-dependent dynamical variables are well defined
at equilibrium. The presence of nonlinear effects or far from equilibrium
dynamics may lead to situations where the FDT does not hold, the aging
process in spin-glass systems being a good example \cite{Parisi,Kauzmann,Rubia,Ricci,Andrea,Exartier,Grigera}. 

In this letter, we present a straightforward proof of the inconsistency
of the FDT for a certain class of superdiffusive processes described
by a generalized Langevin equation (GLE). The surprising result here
is the fact that the violation happens in the linear regime, i.e.
where we expect the validity of the LNESM. The use of the FDT allows
us to classify two classes of superdiffusion. The first class, which
we shall call normal superdiffusion, does obey the FDT; the second
class, which we shall call fast superdiffusion, does not obey the
FDT. The proof is simple and we discuss as well how the diffusive
process leads to an equilibrium. 

Diffusion is one of the simplest processes by which a system reaches
equilibrium. For normal diffusion, the process is so well known that
it may be described by an equilibrium type distribution for the velocity
and position of a particle. However, the strange kinetics of anomalous
diffusion, intensively investigated in the last years \cite{Bouc,Scher,Shlesinger,Srokowski1,Oli2001,Oliveira},
shows surprising results. Consequently, studying anomalous diffusion
seems to be a good way to obtain the conditions of validity for the
FDT.

\section{Violation of the FDT}

We shall start writing the GLE for an operator $A$ in the form \cite{Kubo,Mori,Lee2}

\begin{equation}
\frac{dA(t)}{dt}=-\int _{0}^{t}\Gamma (t-t')A(t')dt'+F(t),\label{1}\end{equation}
 where $F(t)$ is a stochastic noise subject to the conditions $\langle F(t)\rangle =0$,
$\langle F(t)A(0)\rangle =0$ and

\begin{equation}
C_{F}(t)=<F(t)F(0)>=<A^{2}>_{eq}\Gamma (t).\label{2}\end{equation}
 Here $C_{F}(t)$ is the correlation function for $F(t)$ and the
brackets $<>$ indicate ensemble average. Eq. (\ref{2}) is the famous
Kubo FDT and it is quite general\cite{Kubo}. In principle, the presence
of the kernel $\Gamma (t)$ allows us to study a large number of correlated
processes. The main quantity is the correlation function
$C_{A}(t)=<A(t)A(0)>,$ from which we can describe most of the process
of interest.

We may naively expect that, by Eq. (\ref{1}) and Eq. (\ref{2}),
a system will be driven to an equilibrium , i.e. 

\begin{equation}
\lim _{t\rightarrow \infty }<A^{2}(t)>=<A^{2}>_{eq}.\label{2.6}\end{equation}
 We shall see however that this is not always the case for superdiffusive
dynamics. Let us define the variable 

\begin{equation}
y(t)=\int _{0}^{t}A(t')dt',\label{3}\end{equation}
 with asymptotic behavior 

\begin{equation}
\lim _{t\rightarrow \infty }<y^{2}(t)>\sim t^{\alpha }.\label{3.5}\end{equation}
 For normal diffusion $\alpha =1$, we have subdiffusion for $\alpha <1$
and superdiffusion for $\alpha >1$. Notice that if $A(t)$ is the
momentum of a particle with unit mass, $y(t)$ is its position. Using
the Kubo's definition of the diffusion constant, Morgado et al\cite{Oliveira}
obtained a general classification for anomalous diffusion; i.e. using

\begin{equation}
D=\lim _{z\rightarrow 0}\widetilde{C}_{A}(z)=\lim _{z\rightarrow 0}\frac{<A^{2}>_{eq}}{z+\tilde{\Gamma }(z)},\label{4}\end{equation}
 where $\tilde{\Gamma }(z)$ is the Laplace transform of $\Gamma (t)$
they obtained: For a finite value of $\tilde{\Gamma }(0)\neq 0$,  normal
diffusion; for $\tilde{\Gamma }(0)=0,$ superdiffusion; and for $\tilde{\Gamma }(0)=\infty $, 
subdiffusion. 

Now the Laplace transform of Eq. (\ref{1}) suggests a solution of
the form

\begin{equation}
A(t)=\int _{0}^{t}R(t-t')F(t')dt',\label{6}\end{equation}
 where we have set $A(0)=0$, and 

\begin{equation}
\tilde{R}(z)=\frac{1}{z+\tilde{\Gamma }(z)}.\label{7}\end{equation}
 Squaring Eq. (\ref{6}) and taking the ensemble average we obtain 

\begin{equation}
<A^{2}(t)>=2\int _{0}^{t}R(t')\int _{0}^{t'}C_{F}(t'-t'')R(t'')dt''dt'.\label{8}\end{equation}

At this point, it is quite usual to perform numerical calculation
\cite{Srokowski1}. However, we shall show here that very important
results can be obtained analytically. From Eq. (\ref{7}), we can
get a self-consistent equation for $R(t)$ as

\begin{equation}
\frac{dR(t)}{dt}=-\int _{0}^{t}\Gamma (t-t')R(t')dt'.\label{8.5}\end{equation}
Notice from Eq. (\ref{4}) and Eq. (\ref{7}) that $R(t)$ is the
normalized correlation function, i.e. $R(t)=C_{A}(t)/C_{A}(0)$. By
using the FDT Eq. (\ref{2}) and Eq.(\ref{8.5}) we can exactly integrate
Eq. (\ref{8}) and obtain 

\begin{equation}
<A^{2}(t)>=<A^{2}>_{eq}\lambda (t),\label{9}\end{equation}
 where 

\begin{equation}
\lambda (t)=1-R^{2}(t).\label{10}\end{equation}
 Equation. (\ref{2.6}) is satisfied if and only if 

\begin{equation}
\lim _{t\rightarrow \infty }\lambda (t)=\lambda ^{*}=1,\label{10.5}\end{equation}
 or equivalently 

\begin{equation}
\lim _{t\rightarrow \infty }R(t)=\lim _{z\rightarrow 0}z\tilde{R}(z)=\lim _{z\rightarrow 0}\frac{z}{z+\tilde{\Gamma }(z)}=0.\label{11}\end{equation}
 Where in the first step we used the final value theorem\cite{Spiegel}.
Again we shall comment here that if Eq. (\ref{11}) does not hold,
its equivalently Eq. (\ref{2.6}) does not hold as well, and it violates
the FDT Eq. (\ref{2}). Recently Lee\cite{Lee3}, using his recurrence
relation formalism\cite{Lee2}, obtained a condition similar to Eq.  
(\ref{11}) for the ergodicity hypothesis to work. I.e. if the system
keeps some memory after an infinity time, the ergodicity does not
hold. We show here that it is the same condition for the validity
of the FDT.

Equation (\ref{11}) is satisfied for normal and subdiffusion. For
superdiffusion $\lim _{z\rightarrow 0}\tilde{\Gamma }(z)=0$, and
Eq. (\ref{11}) becomes 

\begin{equation}
\lim _{t\rightarrow \infty }R(t)=(1+\lim _{z\rightarrow 0}\frac{\partial \tilde{\Gamma }(z)}{\partial z})^{-1},\label{12}\end{equation}
 There are two distinct limits for Eq. (\ref{12}), which define two
classes of superdiffusion. For the first class, $\lim _{z\rightarrow 0}\frac{\partial \tilde{\Gamma }(z)}{\partial z}=\infty $
and the system obeys the FDT. The second class has $\lim _{z\rightarrow 0}\frac{\partial \tilde{\Gamma }(z)}{\partial z}\neq \infty $
and it does violate the FDT. The first class we shall call normal
superdiffusion (NSD) and the second class fast superdiffusion (FSD). 

Consider now the asymptotic behavior for $\tilde{\Gamma }(z)$as

\begin{equation}
\tilde{\Gamma }(z\rightarrow 0)=az^{\nu }.\label{13}\end{equation}
 Is easy to see that for $\nu <0$ we have subdiffusion, for $\nu =0$
normal diffusion, and for $\nu >0$, superdiffusion. From the above equations we have for $0<\nu <1$
 NSD and, finally, for $\nu \geq 1$ we
have FSD. There is an obvious connection between $\nu $ and $\alpha $,
defined in Eq. (\ref{3.5}), that classify the diffusion. Morgado et al \cite{Oliveira} shown that $\nu =\alpha -1$ and consequently
the FSD starts at $\alpha \geq 2$, i.e., the FDT does not work for
the ballistic motion and beyond. Better, the formalism of GLE + FDT
works for $0<\alpha <2$. We shall discuss later the lower limit $\alpha =0$.

\section{Ballistic Motion }

The ballistic motion is at the limit of the validity of the FDT, to
go beyond it is too dangerous so we shall keep working within $\alpha =2$.
Before we give some example let us make a very important association.
The force $F(t)$ in Eq. (\ref{1}) can be obtained from a thermal
bath composed of harmonic oscillations, consequently accordingly to
Eq. (\ref{2}) the memory can be put as

\begin{equation}
\Gamma (t)=\int \rho (\omega )\cos (\omega t)d\omega ,
\end{equation}
where $\rho $$(\omega )$ is the noise density of states. The same
argument used before for the Laplace transform can be used for for
the Fourier transform, with the simplifying consequence $\tilde{\Gamma }(\omega )=\rho (\omega )$.
This is a great advantage, since noise density of states exist not
only for systems governed by GLE, but for most of the physical systems.
Consequently the MOBH conjecture\cite{Oliveira} reads: \emph{If a
disordered unidimensional system has $\rho $$(\omega )\sim \omega ^{\nu },$
as $\omega \rightarrow 0$ than the diffusion exponent is}

\begin{equation}
\alpha =\nu +1. 
\end{equation}

This conjecture has been observed for the quantum disordered Heisenberg
ferromagnetic chain\cite{Fid1}, and is under discussion for energy
propagation on the harmonic disordered chain\cite{Fid2}. We can choose
now the density of state to produce the diffusion we want. Consider
now the noise density of states 

\begin{equation}
\rho(\omega) = \left\{
  \begin {array}{ll}
 constant & \omega_{1}<\omega<\omega_{2}\\
 0 & otherwise
  \end {array}
  \right.
,          
\end{equation} 
for $\omega _{1}=0$ we have the Debye density of states for a thermal
noise made out of acoustic phonons. Thus, for $\omega _{1}=0$ we
have normal diffusion and for any $\omega _{1}\neq 0$ we have superdiffusion.
This density yields 

\begin{equation}
\Gamma (t)=\beta \left[\frac{\sin (\omega _{2}t)}{t}-\frac{\sin (\omega _{1}t)}{t}\right],\label{14}\end{equation}
 The Laplace transform of Eq. (\ref{14}) gives as $z\rightarrow 0$
$\tilde{\Gamma }(z)$$\sim $$z$, consequently $\nu =1$ and
$\alpha =2$, which is the ballistic limit. If we let $\beta =\omega _{2}/2$
we get $\lambda ^{*}$ as 

\begin{equation}
\lambda ^{*}=1-\left(\frac{2\omega _{1}\omega _{2}}{\omega _{1}+\omega _{2}}\right)^{2}.\label{15}\end{equation}
 Values of $\lambda ^{*}$$\neq $1 show the inconsistency of the
FDT because we start supposing the existence of an equilibrium value
$<A^{2}>_{eq}$ and, after an infinite time, we end up with $<A^{2}>_{eq}\lambda ^{*}$.
Equation (\ref{15}) has a parameter control $\omega _{1}$, which
measures the {}``hole'' in the density of states, and how far we
are from the result predicted by the FDT.

\begin{figure}

\caption{Normalized mean square velocity as a function of time for the memory
given by Eq.(\ref{14}). Here $\beta =\omega _{2}/2$ and $\omega _{2}=0.5$.
Each curve corresponds to a different value of $\omega _{1}$. a)
$\omega _{1}=0$; b) $\omega _{1}=0.25$; c) $\omega _{1}=0.45$.
The horizontal lines correspond to the final average value $\lambda _{s}$.
In agreement with the theoretical prediction, $\lambda _{s}$ decreases
as $\omega _{1}$ grows.}
\end{figure}

Now we select $A(t)=v(t)$, the particle's velocity, so that $<v^{2}(t)>=<v^{2}>_{eq}\lambda (t)$.
We simulate the GLE for a set of $10,000$ particles starting at rest
at the origin and using the memory in Eq. (\ref{14}) with $\omega _{2}=0.5$
and different values of $\omega _{1}$. The results of these simulations
are shown in Fig. 1, where we plot $<v^{2}(t)>$. We used the normalization
$<v^{2}>_{eq}=1$, so that $<v^{2}(t)>=\lambda (t)$. Notice that
$\lambda (t)$ does not reach a stationary value, rather it oscillates
around a final average value $\lambda _{s}$. This value of $\lambda _{s}$
should be compared with $\lambda ^{*}$ obtained from Eq. (\ref{15}). 

\begin{figure}

\caption{$\lambda ^{*}$ as a function of the parameter $w_{1}$. Each dot
corresponds to a value of $\lambda _{s}$ obtained from simulations
like those described in Fig. 1. The line corresponds to the theoretical
prediction given by Eq.(\ref{15}).}
\end{figure}

In Fig. 2 we plot $\lambda ^{*}$ as a function of $\omega _{1}$
as in Eq. (\ref{15}) with fixed $\omega _{2}=0.5$. We also plot the
final average values $\lambda _{s}$ obtained from simulations for
different values of $\omega _{1}$. Notice that as $\omega _{1}$ increases $\lambda ^{*}$ decreases as expected. The agreement between simulations and Eq. (\ref{15}) shows that we can predict the average value 
$\lambda _{s}$, even when the FDT does not work.

   Now we can define $\lambda ^{*}$$=T^{*}/T$,
where $T^{*}$ is an effective temperature for the system. Effective
temperatures different of the expected temperature $T$, or $\lambda ^{*}$$\neq 1,$
are found in spin glasses where the FDT does not work\cite{Kauzmann,Rubia,Ricci,Andrea}.
The first observation of such phenomena was reported Kauzmann\cite{Kauzmann}.
He noticed that if the entropy of a supercooled liquid is extrapolated
below the glass temperature $T_{g}$, it becomes equal to the crystal
temperature $T_{c}>0$, and in some cases even $T_{c}<0$. To avoid
this paradox he suggested the existence in the supercooled liquid
phase of effective spinodal temperature $T_{sp}<$$T_{c}$. In a recent
work Rubi et al\cite{Rubia} investigate the violation of the FDT,
using a Fokker-Planck approach. They found temperatures $T^{*}$ which
are greater and small than $T.$ Ricci-Tersenghi et al\cite{Ricci}
and Cavagna et al\cite{Andrea} performed single-spin-flip Monte Carlo
simulations in square lattices with frustration and they obtained
effective temperatures $T^{*}\neq T$. Methods for measuring those
effective temperatures are discussed as well in the literature\cite{Exartier}.

An oscillatory behavior similar to that found in Fig. 1 was observed
by Srokowski\cite{Srokowski1,Oli2001}, in his simulations using GLE.
However, the kind of motion he studied is a subdiffusive motion. Using
his memory we get\cite{Oli2001} $\tilde{\Gamma }(z)=\beta [{1-\exp (\varepsilon z)}/(\varepsilon z)+E_{1}(\varepsilon z)]$,
here $\beta $ is a constant, $\varepsilon $ a small number, and
$E_{1}$ is the exponential integral. This result gives $\lim _{z\rightarrow 0}\tilde{\Gamma }(z)\sim z^{-1}$,
from the above arguments we get $\nu =-1$ and $\alpha =0$.
We believe we can explain this strange results. Let us consider a constant
memory of the form $\Gamma (t)=K,$ if $A(t)$ in Eq. (\ref{1}) is
the velocity, than the memory term yields $-Ky$, where
$y$ is the position. Consequently the particle is bound to the origin
by a harmonic spring and has no diffusive behavior. Indeed the Laplace
transform gives $\tilde{\Gamma }(z)=K/z,$ with $\nu =-1$ and
$\alpha =0$, the same as the result we obtain from Srokowski memory.
 We shall notice
that $\alpha =0$ does not meaning that the motion is a harmonic type
motion, rather it means it belongs to same class, the diffusion behavior
of Srokowski is not a power low as in Eq. (\ref{3}) probably it is
slower than a power low, such as, for example, a logarithmic behavior. 

For $\alpha >2$, the FSD cannot be described by the methods we used
here. Once the FDT does not work, the GLE and the FDT together predict
strange results such as a null dispersion for the dynamical variable,
i.e. $<A^{2}(t\rightarrow \infty )>=0.$ Moreover, the exponent $\alpha $
can be put as $\alpha =2/D_{F}$, where $D_{F}$ is the fractal dimension
\cite{Ord}. Consequently $\alpha >2$ leads to $D_{F}<1$, which
is not a full curve, but a set of points such as the Cantor set, and
cannot represent a classical trajectory.

\section{Conclusion}

We discussed the stationary behavior for the mean square value of
a dynamical variable $A(t)$ and as well the mean square displacement
of the quantity' $y(t)=\int _{0}^{t}A(t')dt'$. The assymptotic behavior
of $<y^{2}(t)>\sim t^{\alpha }$ as $t\rightarrow \infty $, can be
explained for $0<\alpha <2.$ We show the the superdiffusive motion
must be classified in normal superdiffusive (NSD), for $1<\alpha <2,$ and
fast superdiffusive (FSD), for $\alpha \geq 2$. The FSD motion shows
an inconsistency between the GLE and the FDT. This
kind of superdiffusion with $\alpha \geq 2$ is common in hydrodynamical
processes. It is not surprising that these processes will be far from
equilibrium and violate the FDT. We pointed out here how it happens
and precisely where the FDT breaks down. As we have already mentioned,
spin glasses seem to be a rich field for studying these phenomena.
Indeed experimental \cite{Grigera} and theoretical works \cite{Parisi,Ricci,Rubia}
have been reported in this area, confirming the violation of the FDT.
As well, the effective temperature found in noncrystaline material
is connected here with the FSD and the violation of the FDT. It would
be very helpful if the exponent $\alpha $ for those diffusive processes
could be measured. Another related phenomena are anomalous reaction rate\cite{Oliveira2} and chaos synchronization\cite{Oliveira3}, which we expect to discuss
soon. Although anomalous diffusion remains as a surprising phenomena,
we hope that this work will help in the centennial effort to understand
diffusion and the relation between fluctuation and dissipation. A
generalization of the FDT to include the FSD is necessary, what will
require a deeper understanding of systems far from equilibrium.

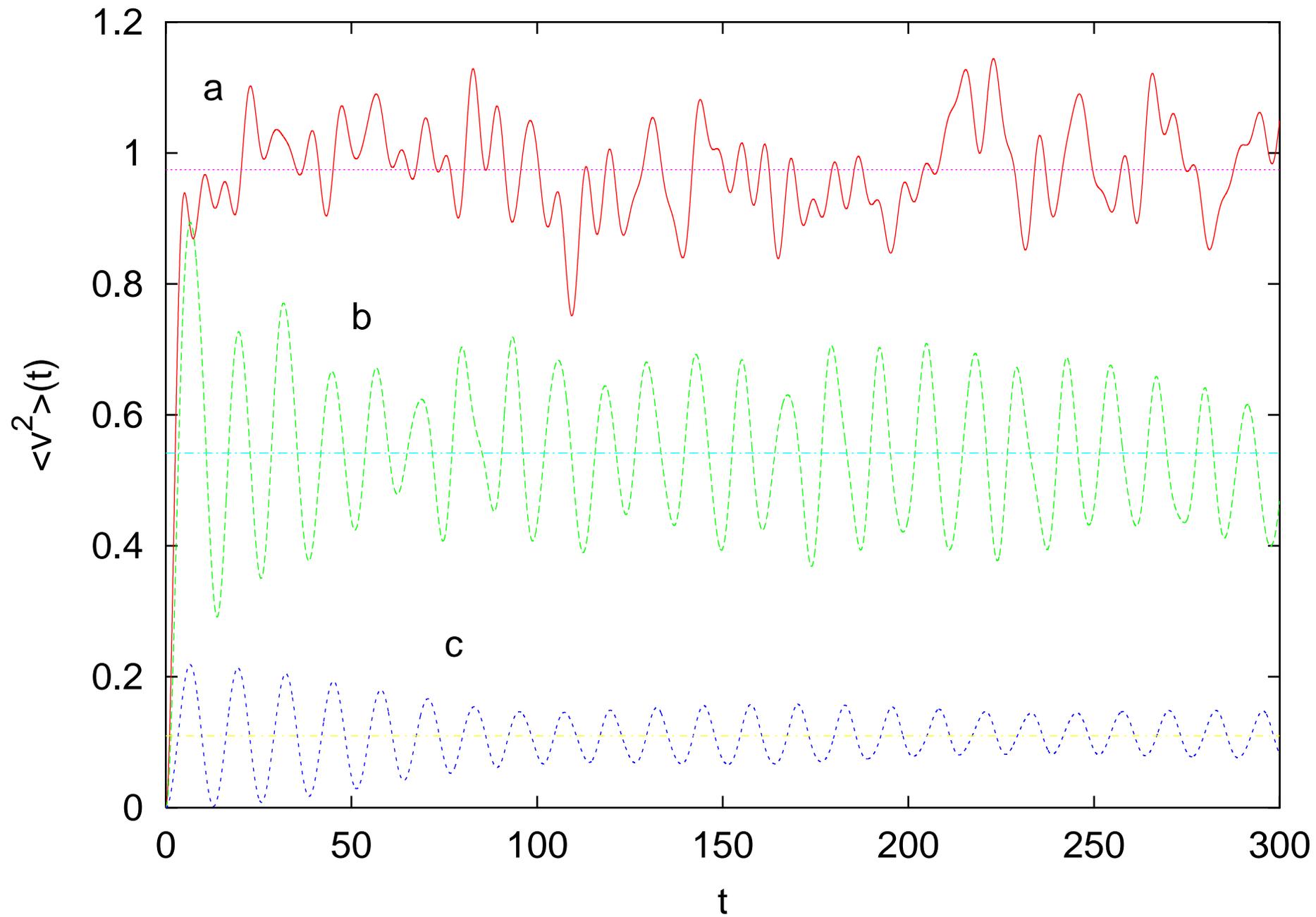
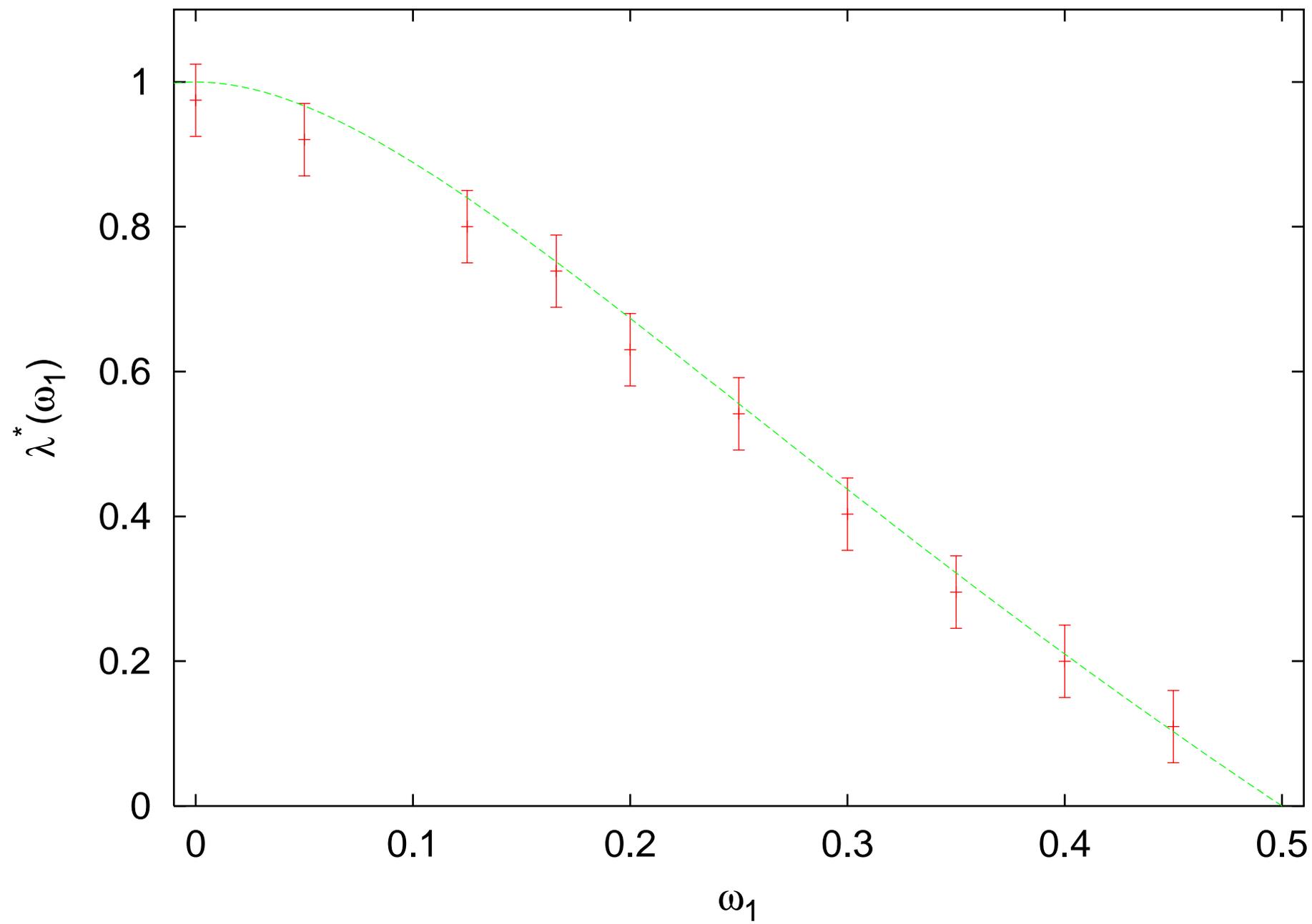
\end{document}